\documentclass[9pt,conference]{IEEEtran}
\IEEEoverridecommandlockouts
\usepackage{cite}
\usepackage{amsmath,amssymb,amsfonts}
\usepackage{algorithmic}
\usepackage{graphicx}
\usepackage{textcomp}
\usepackage{xcolor}
\usepackage{soul}
\usepackage{xspace}
\usepackage{braket}
\usepackage{url}
\usepackage{authblk}
\makeatletter
\renewcommand\AB@affilsepx{, \protect\Affilfont}
\makeatother

\newcommand{\ignore}[1]{}
\def\BibTeX{{\rm B\kern-.05em{\sc i\kern-.025em b}\kern-.08em
    T\kern-.1667em\lower.7ex\hbox{E}\kern-.125emX}}

\usepackage{booktabs}

\usepackage{tikz}
\newcommand*\circled[1]{\tikz[baseline=(char.base)]{
                \node[shape=circle,fill,inner sep=0.5pt] (char) {\textcolor{white}{#1}};}}

\DeclareMathOperator{\inp}{\text{in}}
\DeclareMathOperator{\out}{\text{out}}
\DeclareMathOperator{\adc}{\text{ADC}}
\DeclareMathOperator{\dac}{\text{DAC}}
\DeclareMathOperator{\xhp}{\overrightarrow{X}_{\text{HP}}}
\DeclareMathOperator{\whp}{\overrightarrow{W}_{\text{HP}}}
\DeclareMathOperator{\xlp}{\overrightarrow{X}_{\text{LP}}}
\DeclareMathOperator{\wlp}{\overrightarrow{W}_{\text{LP}}}
\DeclareMathOperator{\xlpi}{\overrightarrow{X}_{\text{LP},i}}
\DeclareMathOperator{\wlpi}{\overrightarrow{W}_{\text{LP},i}}

\begin{document}

\title{Leveraging Residue Number System for Designing High-Precision Analog Deep Neural Network Accelerators}


\author[1]{Cansu Demirkiran}
\author[1]{Rashmi Agrawal}
\author[2]{Vijay Janapa Reddi}
\author[3]{Darius Bunandar}
\author[1]{Ajay Joshi}

\affil[1]{Boston University}
\affil[2]{Harvard University}
\affil[3]{Lightmatter}


\maketitle

\begin{abstract}
Achieving high accuracy, while maintaining good energy efficiency, in analog DNN accelerators is challenging as high-precision data converters are expensive. 
In this paper, we overcome this challenge by using the residue number system (RNS) to compose high-precision operations from multiple low-precision operations. 
This enables us to eliminate the information loss caused by the limited precision of the ADCs.
Our study shows that RNS can achieve $99\%$ FP32 accuracy for state-of-the-art DNN inference using data converters with only $6$-bit precision.
We propose using redundant RNS to achieve a fault-tolerant analog accelerator.
In addition, we show that RNS can reduce the energy consumption of the data converters within an analog accelerator by several orders of magnitude compared to a regular fixed-point approach.  
\end{abstract}

\section{Introduction}
\label{sec:introduction}

Deep Neural Networks (DNNs) are commonly used today in a variety of applications including financial, healthcare, and transportation.
The pervasive usage of these DNN models, whose sizes are continuously increasing, forces us to use more compute, communication, and memory resources. 
Unfortunately, with Moore's Law and Dennard Scaling slowing down~\cite{end-of-moores-law}, we can no longer rely on technology scaling.
As a result, these ever-growing DNNs come with higher and higher costs in terms of energy, time, money, and environmental impact.

For performing efficient DNN inference, several works have explored analog accelerator designs.
These analog designs accelerate general matrix-matrix multiplication (GEMM) operations that make up more than $90\%$ of the operations in DNN inference.
The prior art have explored several analog technologies such as photonic cores~\cite{CoherentNanophotonic2017,  11-tops, dnnara, pixel-2020, albireo-2021}, , resistive arrays~\cite{yao2020fully, prime-2016, isaac, 1t1m-2016, tang-2017}, switched capacitor arrays~\cite{bankman-2015, bankman-sc16}, PCM~\cite{feldmann2021parallel}, STT-RAM~\cite{8241447, 9097177}, etc., to enable highly parallel, fast, and efficient matrix-vector multiplications (MVMs) in the analog domain. 


The success of this analog approach is, however, constrained by the limited precision of the digital-to-analog and analog-to-digital data converters (i.e., DACs and ADCs) which typically dominate the energy consumption in analog accelerators~\cite{isaac, rekhi-2019, Kim:18}.
Concretely, during MVM, a dot product between a $b_w$-bit signed weight vector and a $b_{in}$-bit signed input vector---each with $h$ elements---produces an output scalar with $b_{\out}$ bits of information, where $b_{\out} = b_{\inp} + b_w + \log_2(h)-1$.
Therefore, an ADC with precision greater than $b_{\out}$ is required to guarantee no loss of information when capturing the output.
Unfortunately, the energy consumption of ADCs increases exponentially with the effective number of bits (ENOB) (roughly 4$\times$ increase for each additional output bit)~\cite{murmann-mixed-signal}. 
So for energy-efficient analog accelerator designs, it is typical to use ADCs with a lower precision ($b_{\adc} < b_{\out}$) and only read the most significant bits (MSBs) of the output~\cite{rekhi-2019}.

\begin{figure}[t]
\centering
\includegraphics[width=\linewidth]{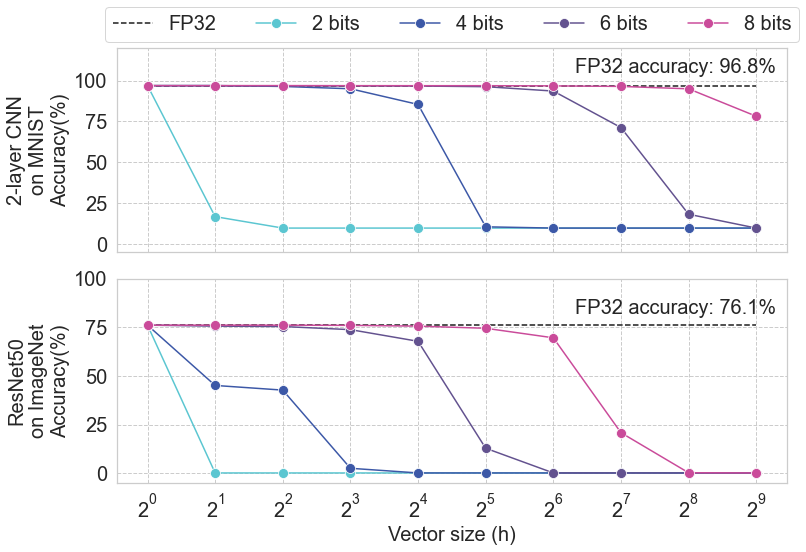}
\caption{Accuracy of a two-layer CNN classifying handwritten digits from the MNIST dataset and ResNet50 classifying images from the ImageNet dataset evaluated in an analog core with varying precision $b$ and vector sizes $h$. For all evaluations, $b$-bit precision means we set $b_{\inp}=b_w=b_{\adc}=b$.}
\label{fig:acc-tile-size}
\end{figure}

Rekhi et al.~\cite{rekhi-2019} showed that while this limited precision does not cause a significant accuracy drop in small networks with small datasets, it can drastically degrade the accuracy in larger networks.
To better understand this, we analyzed (see Fig.~\ref{fig:acc-tile-size}) how the precision of DACs (i.e., $b_{\inp}$ and $b_w$) and ADCs (i.e., $b_{\adc}$), and the number of elements within a single vector (i.e, $h$) affect the accuracy in two tasks: (1) a two-layer convolutional neural network (CNN) for classifying the MNIST dataset~\cite{mnist}: a simpler task with only 10 classes, and (2) ResNet50~\cite{resnet}, a CNN with 50 layers, for classifying the ImageNet dataset~\cite{imagenet}: a more difficult task with 1000 classes.
As the vector size $h$ increases, $b_{\out}$ increases, therefore, higher precision is needed to maintain the accuracy.
At all bit precisions, the accuracy of ResNet50 degrades at smaller values of $h$ than the two-layer CNN.
Essentially, to efficiently execute large DNNs using analog accelerators while maintaining high accuracy, we need a cheaper way to increase the computation precision than simply increasing the bit precision of the data converters.

\begin{figure*}[ht]
\centering
\includegraphics[width=0.9\linewidth]{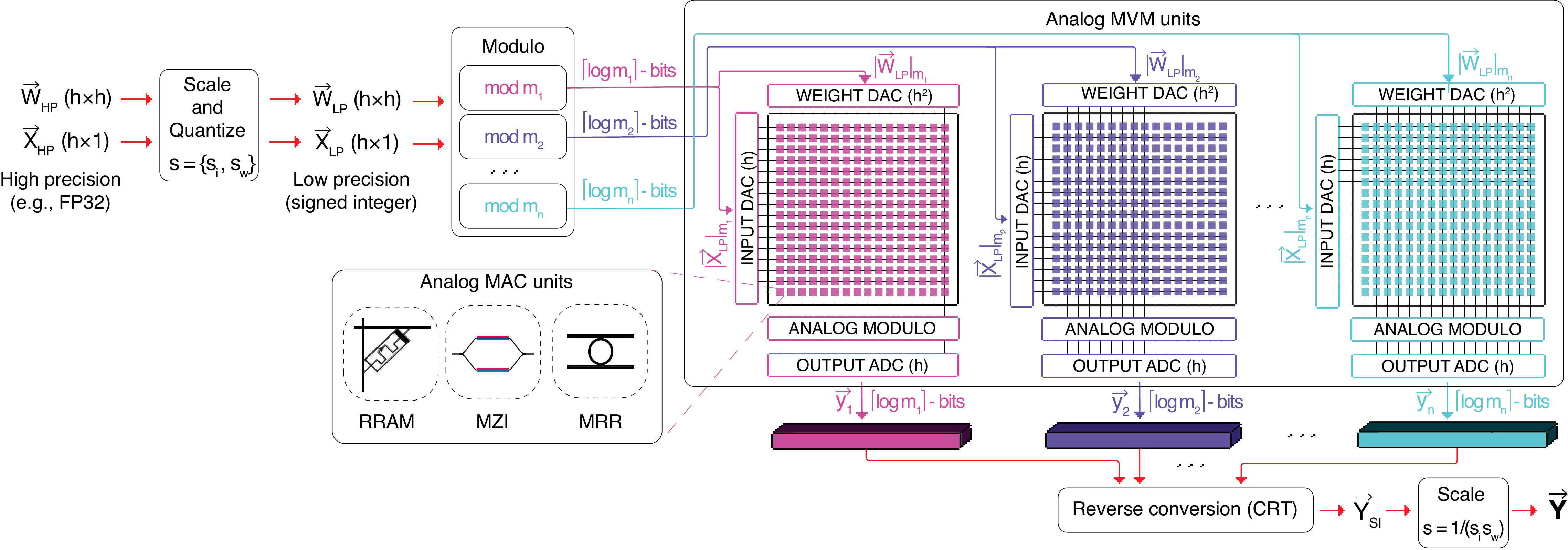}
\caption{Dataflow for the RNS-based analog GEMM operation for a moduli set  $m = \{m_1, \dots, m_{n}\}$. The $n$ $h\times h$ analog MVM units are represented as generic blocks. The MAC units can be resistive memory (RRAM) arrays, photonic GEMM cores, etc. The dataflow is agnostic to the technology. }
\label{fig:rns-core}
\end{figure*}

In this paper, we propose to leverage the residue number system (RNS) to design analog accelerators, to enable the use of low-precision ADCs while performing high-precision arithmetic.
We take advantage of the fact that RNS can represent high-precision values using low-precision integer residues for a chosen set of moduli. 
As RNS is closed under multiplication and addition, a full dot-product can be performed without switching between numeral systems.
The modulo operation in the RNS ensures that the bit width of the residues does not grow to be larger than the bit width of the chosen moduli.
This keeps the output residues at the same bit precision as the input residues after a dot product.
Therefore, the low-precision output residues can be captured by low-precision ADCs without any information loss and eventually be used to generate a high-precision GEMM output.

Our contributions in this work are as follows:
\circled {1} We present a dataflow for a generic analog RNS-based accelerator.
We evaluate the effectiveness of the proposed RNS approach by comparing the accuracy of an RNS-based analog hardware against a regular fixed-point-based analog hardware using state-of-the-art MLPerf (datacenters) DNN benchmarks~\cite{ml-perf-2020}.  
\circled {2} To improve the fault tolerance of the analog accelerator, we incorporate the Redundant RNS (RRNS) error-correcting code~\cite{rrns-2015} in the analog RNS-based accelerator.
We use redundant residues to detect and correct errors at the output of a dot product.
We evaluate the fault tolerance of the error-corrected analog accelerator by injecting noise and investigating its effects on the accuracy of the MLPerf benchmarks.
\circled {3} We investigate the energy efficiency advantage of using our RNS approach over regular fixed-point number system in analog accelerators.
Our results show that, thanks to the use of low-precision data converters enabled by our RNS-based approach, the energy consumed by data converters (that dominate the energy consumption in analog accelerators) can be reduced by multiple orders of magnitude compared to the same precision regular analog approach.

\section{Related Work}
\label{sec:rel-wk}




Prior works have investigated several analog technologies such as photonic cores~\cite{CoherentNanophotonic2017,  11-tops, dnnara, pixel-2020, albireo-2021}, resistive arrays~\cite{yao2020fully, prime-2016, isaac, 1t1m-2016, tang-2017}, switched capacitor arrays~\cite{bankman-2015, bankman-sc16}, PCM cells~\cite{feldmann2021parallel}, STT-RAM\cite{8241447, 9097177}, etc., for DNN acceleration. 
Many of these analog DNN accelerator works~\cite{CoherentNanophotonic2017, 11-tops,yao2020fully} have mainly reported accuracy results only for easy tasks such as classification of small datasets (e.g., MNIST, CIFAR) that work well with very low precision.
A few other works~\cite{albireo-2021, pixel-2020} used larger and more recent networks in their evaluation but solely focused on hardware performance and architectural design without reporting accuracy with the assumption that 8-bit DACs and ADCs provide adequate precision for high-accuracy DNN inference. 
However, as shown in Figure~\ref{fig:acc-tile-size}, this assumption may not be valid for all DNNs and hardware configurations.
\emph{In our work, we focus on state-of-the-art networks instead of small tasks that are practically obsolete.}

RNS has been used in the analog domain for building optical adders and multipliers for reducing the optical critical path~\cite{Peng:18} and increasing efficiency in DNNs~\cite{dnnara}. 
However, this requires a number of optical devices that increases quadratically with the modulus value---degrading their efficiency for large moduli.
\emph{In our work, the number of analog devices is independent of the size of the moduli.}

There also exist digital DNN accelerators~\cite{rns-net, res-dnn} that use RNS for energy-efficient computation.
These works show that it is possible to stay in the RNS domain (without reverting back to binary or decimal) for the entire inference. 
However, this approach requires overflow detection and scaling and also uses approximations for non-linear operations.
Most state-of-the-art DNNs today comprise a wide variety of non-linear operations. 
Using approximations and fixed-point-only arithmetic for these operations can severely degrade the accuracy. 
Therefore, \emph{in our work, we use RNS only for MVM operations and switch back to floating-point arithmetic for non-linear operations.}

Bit-partitioned arithmetic~\cite{bit-partition-mixed-signal} is another way to build high-precision results from low-precision operations and eliminate the need for costly high-precision ADCs. 
Shafiee et al. \cite{isaac} use an iterative process where 16-bit inputs are represented as 16 1-bit values to eliminate the DACs and reduce the required precision of ADCs.
However, their iterative approach takes more than one cycle to compute the dot products.
In addition, both the RNS and the bit-partitioning approaches are susceptible to noise as small errors in the residues/partitions grow larger during output reconstruction.
Thus, error correction methods are required for such designs.  
\emph{So in our work, we provide a complete fault-tolerant dataflow that uses redundant RNS to detect and correct errors that are introduced by the noisy behavior of analog designs.} 

\section{RNS-based Analog Core}
\label{sec:rns-core}

\begin{table*}[t]
    \centering
    \vspace{-0.1in}
    \caption{
    Comparison of RNS-based and regular fixed-point analog cores for different input and weight precision and $h$ = 128.
    }
    \label{table:moduli-sets}
    \begin{tabular}{cccccccccc}
        \toprule
        &\multicolumn{5}{c}{\textbf{RNS-based Analog Core (This work)}} & \multicolumn{4}{c}{\textbf{Regular Fixed-Point Analog Core with $b_{\adc} < b_{\out}$}} \\
        \cmidrule(lr){2-6}
        \cmidrule(lr){7-10}
        $b_{\inp}$, $b_w$ & $b_{\dac}$ & ${\log}_2(\mathcal{M}$ & $b_{\adc}$ &  Example Moduli Set ($\mathcal{M})$ & RNS Range ($M$) & $b_{\dac}$ & $b_{\out}$ & $b_{\adc}$  & Num. of Lost Bits\\ 
        \cmidrule(lr){1-1}
        \cmidrule(lr){2-2}
        \cmidrule(lr){3-3}
        \cmidrule(lr){4-4}
        \cmidrule(lr){5-5}
        \cmidrule(lr){6-6}
        \cmidrule(lr){7-7}
        \cmidrule(lr){8-8}
        \cmidrule(lr){9-9}  
        \cmidrule(lr){10-10}
        4 & 4 &4 & 4 & $\{15, 14, 13, 11\}$ & $\simeq 2^{15} - 1$ &4& 14 & 4 & 10  \\
        5 & 5 & 5 &5 & $\{31, 29, 28, 27\}$ & $\simeq 2^{19} - 1$ &5& 16 & 5  & 11 \\
        6 & 6 & 6 &6 & $\{63, 62, 61, 59\}$ & $\simeq 2^{24} - 1$ &6& 18 & 6  & 12 \\
        7 & 7 & 7 &7 & $\{127, 126, 125\}$ & $\simeq 2^{21} - 1$ &7& 20 & 7 & 13 \\
        8 & 8 & 8 & 8 &$\{255, 254, 253\}$ & $\simeq 2^{24} - 1$ &8& 22 & 8 & 14 \\
        \bottomrule
    \end{tabular}
\end{table*}

\subsection{RNS Basics}
The RNS represents an integer as a set of smaller integer residues.
These residues are calculated by performing a modulo operation on the said integer using a selected set of $n$ \emph{co-prime} moduli. 
Let $A$ be an integer. 
$A$ can be represented in the RNS with $n$ residues as $\{a_1, \dots, a_{n}\}$ for a set of moduli $\mathcal{M} = \{m_1, \dots, m_{n}\}$ where $a_i = |A|_{m_i} \equiv A \mod m_i$ for $i \in \{1\dots n\}$\footnote{Hereinafter, we refer to the integer $A$ as the \emph{standard representation}, while we refer to the integers $\{ a_1, \dots, a_n\}$ simply as the residues.}
. 
$A$ can be uniquely reconstructed using the Chinese Remainder Theorem (CRT):
\begin{equation}
A = \sum_{i=1}^n|a_i M_i T_i|_M,
    \label{eq:crt}
\end{equation}
if all the moduli are relatively prime and $A$ is within the range $[0, M)$ where $M= \prod_i m_i$. 
Here, $M_i = M/m_i$ and $T_i$ is the multiplicative inverse of $M_i$, i.e., $\left| M_i T_i \right|_{m_i} \equiv 1$. 
The RNS is closed under addition and multiplication operations, thus enabling dot products and GEMM operations in the RNS space.

\subsection{DNN Inference Using RNS}
A DNN consists of a sequence of $L$ layers.
The input $\vec{X}$ to the $\ell$-th layer of a DNN inference is the output generated by the previous $(\ell-1)$-th layer:
\begin{equation}
    \vec{X}_k^{(\ell)} = f^{(\ell-1)} \left(\sum_j W_{jk}^{(\ell-1)} \vec{X}_j^{(\ell-1)} \right),
    \label{eq:nn}
\end{equation}
where $W \cdot \vec{X}$ is an MVM and $f(\cdot)$ is an element-wise nonlinear function.

In both digital and analog designs, the precision of MVM is limited by the precision of the inputs and weights, as well as the amount of information kept at the output.
In digital accelerators that use low precision, when performing an MVM, costly multiplications are performed with low-precision datatypes (e.g., INT8) and accumulations are performed with high-precision datatypes (e.g., INT32/FP32), as accumulation is cheaper. 
The final result of the MVM can then be stored as a low-precision number. 
In contrast, in analog accelerators, the inputs and weights are both multiplied and accumulated with low-precision datatypes.
This is because the fixed size of the analog compute array typically does not let the whole GEMM operation (a full DNN layer) be performed at once and each MVM produces a partial output vector which needs to be captured by the ADCs before being stored in a digital memory unit, i.e., SRAM or DRAM, and being accumulated with the rest of the partial output vectors.
Here, the ADC precision directly determines how precisely we can capture these partial outputs.
While using low-precision ADCs will lead to a higher information loss compared to digital hardware that can severely impact the accuracy of DNN, using a high-precision ADC is not feasible as it is power-hungry.
However, this extra loss of information due to using low-precision ADCs can be eliminated by using RNS.
Using RNS, Eq.~\eqref{eq:nn} can be rewritten as:
\begin{equation}
    \vec{X}_k^{(\ell)} = f^{(\ell-1)} \Bigg(\text{CRT}\bigg(\Big|\sum_j
    \big|W_{jk}^{(\ell-1)}\big|_{\mathcal{M}}
    \big|\vec{X}_{j}^{(\ell-1)}\big|_{\mathcal{M}}
     \Big|_{\mathcal{M}}\bigg)\Bigg).
\end{equation}
Here $\mathcal{M} = \{m_1, \dots, m_{n}\}$ represents the set of moduli. 
The moduli must be chosen to ensure that the outputs of the MVM operations are smaller than $M=\prod_i m_i$, which means we need
\begin{equation}
    \log_2 M \geq b_{\out} = b_{\inp} + b_w + \log_2(h)-1,
    \label{eq:rns_bit_inequality}
\end{equation}
for a dot-product between two vectors with $h$-elements.
This constraint prevents overflow in the computation. 
Operations for different residues can be performed independently and in parallel without any carry propagation. 

\begin{figure}[t]
\centering
\includegraphics[width=\linewidth]{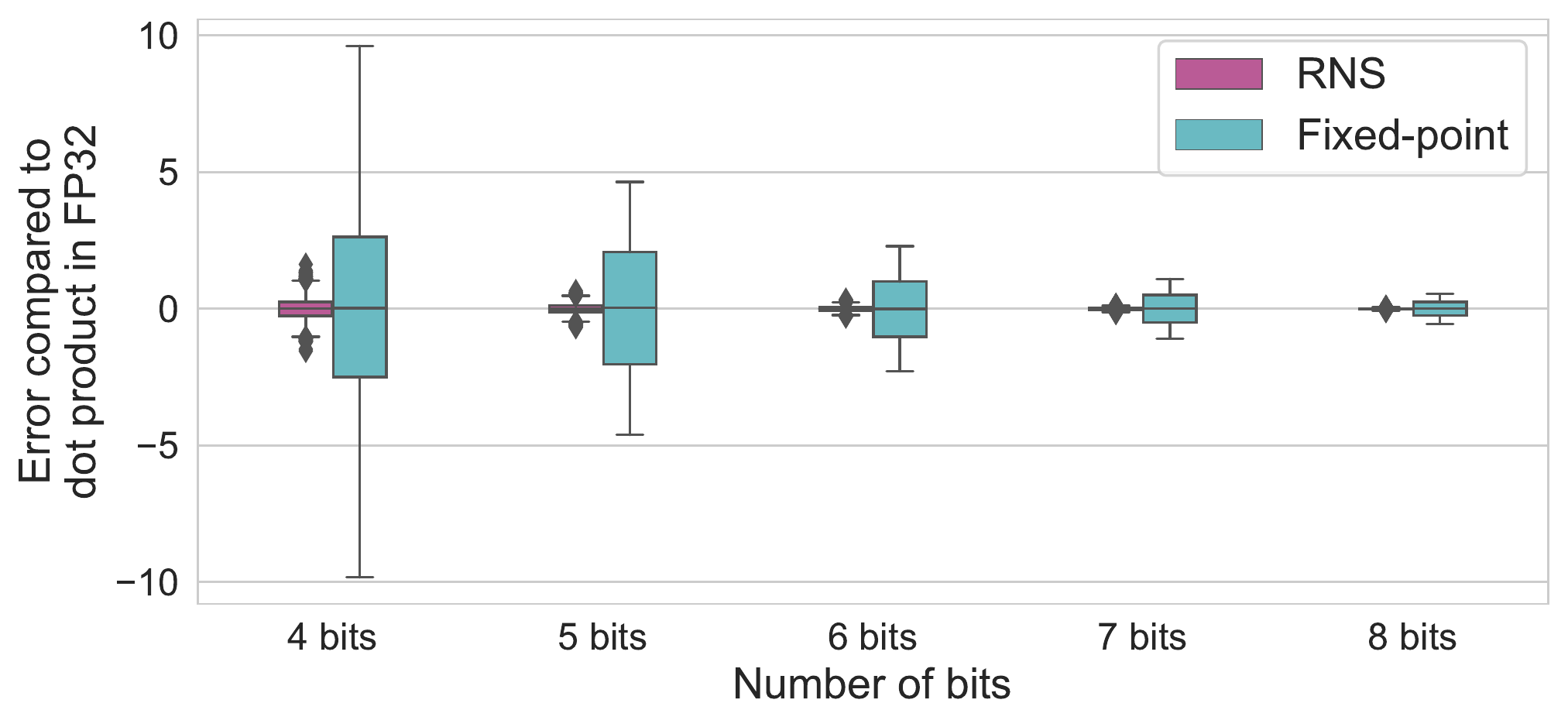}

\caption{Distribution of the error in the dot product performed with regular fixed-point analog core and RNS-based analog core. The error is the difference between the dot product calculated in FP32 (FP32 being the ground truth), and dot product calculated in fixed-point and RNS. Both fixed-point and RNS-based analog cores use the same input, weight, DAC, and ADC precision which varies from 4 to 8 in the plot. In the fixed-point core, only the 4-8 MSBs of the output are kept. The RNS-based core uses moduli within the same number of bits of range (see Table~\ref{table:moduli-sets}). The results are shown for randomly generated 10,000 vector pairs for each case.}

\label{fig:dot-prod-err}
\end{figure}

Fig.~\ref{fig:rns-core} shows the dataflow for the RNS-based analog hardware when performing MVM as part of the DNN inference.
Here, $\xhp$  represents a high-precision (e.g., FP32) $h\times 1$ input vector and $\whp$ represents a high-precision $h\times h$ weight matrix.\footnote{For inputs and weights with dimensions larger than $h$, one can use standard tiling methods.}
The input vector and the weight matrix have to be mapped to integers to work with RNS. 
To reduce the quantization effects, we scale both inputs and weights before quantization.  
We scale $\xhp$ by dividing each element of the vector by $s_{\inp} = \text{max}(|\xhp|)$.
For the $h\times h$ weight matrix, we scale each row separately. 
The scaling factors for the $h$ rows form a vector $\vec{s}_w = [s_w[1], s_w[2], \dots, s_w[h]]=[ \text{max}(|\whp[1]|), \text{max}(|\whp[2]|), \dots,  \text{max}(|\whp[h]|)]$. 
In total, there are $h+1$ floating-point scaling factors: $h$ for the weight matrix and one for the input vector.
These scaled elements are then quantized and mapped to (symmetric) signed integers between $[-2^{b-1}-1,2^{b-1}-1]$ with $b=b_{\inp}$ and $b=b_w$ to obtain the low-precision input vector ($\xlp$) and weight matrix ($\wlp$), respectively. 
We then perform the modulo operation on each element of the $\xlp$ and $\wlp$ with respect to each of the $n$ moduli.
This gives us $n$ pairs of $\xlpi$ and $\wlpi$ containing the corresponding residues as its elements, where $i \in \{1, \dots, n \}$.

\begin{figure*}[ht]
\centering
\includegraphics[width=0.9\textwidth]{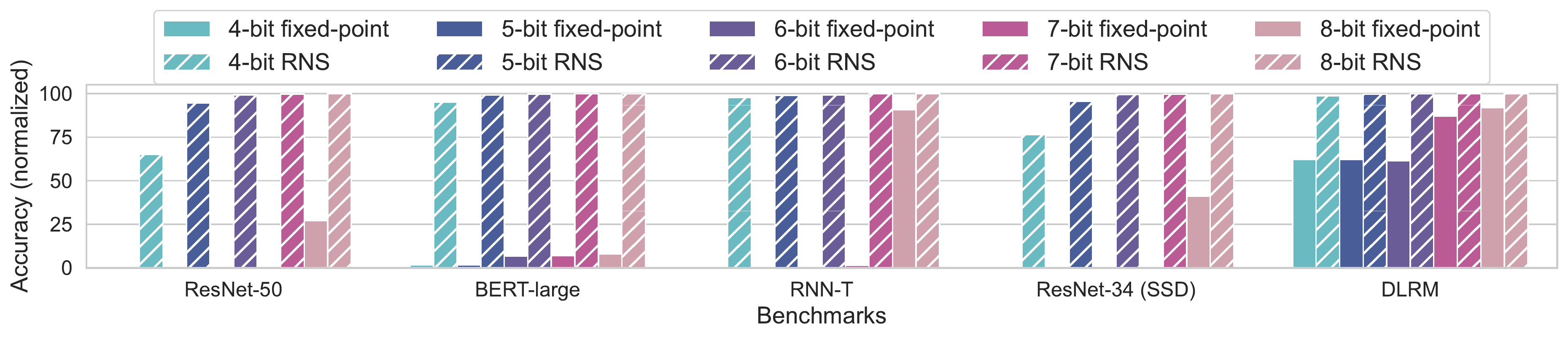}
\caption{Accuracy of regular fixed-point and RNS-based cores on MLPerf datacenter benchmarks. The accuracy numbers are normalized to the FP32 accuracy. Both $b$-bit fixed-point cores and $b$-bit RNS-based cores use $b$-bit input and weight DACs and $b$-bit output ADCs. The RNS-based cores use $b$-bit moduli to represent numbers (see Table~\ref{table:moduli-sets}). All experiments use FP32 as the intermediate datatype for non-GEMM operations.}
\label{fig:acc-no-error}
\end{figure*}

In the RNS-based analog core, we have a dedicated analog MVM unit for each modulus.
Each analog MVM unit has a set of DACs for converting the associated input vector and weight matrix into the analog domain.
An analog MVM is performed (i.e., $h$ dot products in parallel) between the input vector and the weight matrix, followed by a modulo operation on each output residue vector in the analog domain.
Thanks to the modulo operation, the output residues are reduced back to $[0,m_i)$ range. 
These output residues are captured by ADCs. 
An ENOB of $\lceil\log_2{m_i}\rceil$ (instead of the larger $b_{\out}$-bits) is adequate for both DACs and ADCs to perform input and output conversions without any information loss. 

The output residues are then converted back to the standard representation in the digital domain via CRT (Eq.~\eqref{eq:crt}) to generate the signed integer output vector, i.e., $\overrightarrow{Y}_\text{SI}$.
The elements of the output vector are then scaled back up using the scaling factors calculated before the MVM operation, i.e., $Y[k] = Y_{\text{SI}}[k]\cdot s_{\inp}\cdot s_{w}[k]$.
The non-linear function $f$ (e.g., ReLU, sigmoid, etc.) is performed digitally on the MVM output using floating-point datatypes (e.g., FP32). 

\subsection{Precision in the RNS-based Analog Core}
\label{sec:precision}

To obtain a desired output precision, we need to find a set of moduli that meets Eq.~\eqref{eq:rns_bit_inequality}. 
The RNS range ($M$) depends on both the number of moduli ($n$) and the values of these moduli.
The number of moduli determines the number of operations that we need to perform in the RNS-based core along with the required number of MVM units.
Similarly, the bit width of these moduli determine the required precision of the data converters.
Both of these in turn determine the efficiency of the RNS-based analog core and should be chosen carefully to find a balance between the energy efficiency and $M$.


Table~\ref{table:moduli-sets} shows the comparison of our RNS-based analog core with example moduli sets and a regular fixed-point analog core\footnote{Here both cores use fixed-point numbers to represent data. We use the term `regular fixed-point core' for referring to a typical analog core that uses fixed-point numbers for computation and do not use RNS, whereas the RNS-based core uses integers as residues and performs computations in the RNS space.}.
For the RNS-based core, we picked ADC and DAC precision to be the same as the precision of the input vector and the weight matrix, i.e., $b = b_{\inp} = b_{w} = b_{\adc} = b_{\dac} = \lceil \log_2m_i\rceil $.
The moduli sets shown in Table~\ref{table:moduli-sets} are created by using the minimum number of moduli that guarantees Eq.~\eqref{eq:rns_bit_inequality} for $h = 128$ while keeping the moduli under the chosen bit width $b$ \footnote{We choose $h$ to be 128 considering the common layer sizes in the evaluated MLPerf benchmarks. 
The chosen $h$ provides high throughput with a high utilization of the GEMM core.
}.
In this case, for $n$ moduli, $M$ is $\approx n \cdot b$ bits.
It should be noted that the values of $h$ and $\mathcal{M}$ are representative values and our methods can be generalized for any $h$ and $\mathcal{M}$. 
In a regular fixed-point analog core, similar to the RNS-based core, we set $b = b_{\inp} = b_{w} = b_\text{DAC}= b_\text{ADC}$.  
However, the bit-precision required to represent the output produced by the dot product ($b_{\out}$) is much larger than $b_{\adc}$. 
Therefore, $b_{\out}- b_{\adc} $ bits of information (from LSB and up) are lost after every MVM operation.

Fig.~\ref{fig:dot-prod-err} reports the absolute errors (FP32 being the ground truth) observed when performing dot-products (sub-operations in MVMs) with the RNS-based and fixed-point analog cores.
Both cores use the configurations described in Table~\ref{table:moduli-sets} for the example vector size $h = 128$.
In the regular fixed-point core, information loss due to the dropped bits causes a 9--15$\times$ larger error compared to the RNS-based dot product with the same input and weight bit precision.

We next compare the performance of an RNS-based analog core against a fixed-point analog core for different MLPerf benchmarks as shown in Fig.~\ref{fig:acc-no-error}.
We report accuracy normalized to the accuracy of a FP32 hardware.
Our results show that the RNS approach significantly ameliorates the accuracy drop caused by the low-precision ADCs in the fixed-point analog core for all the benchmark networks.
We observe that it is possible to achieve ${\geq}99\%$ of FP32 accuracy (this cut-off is defined in the MLPerf benchmark~\cite{ml-perf-2020}) for all MLPerf networks when using residues with as low as $6$ bits.

\begin{figure*}[ht]
\centering
\includegraphics[width=\textwidth]{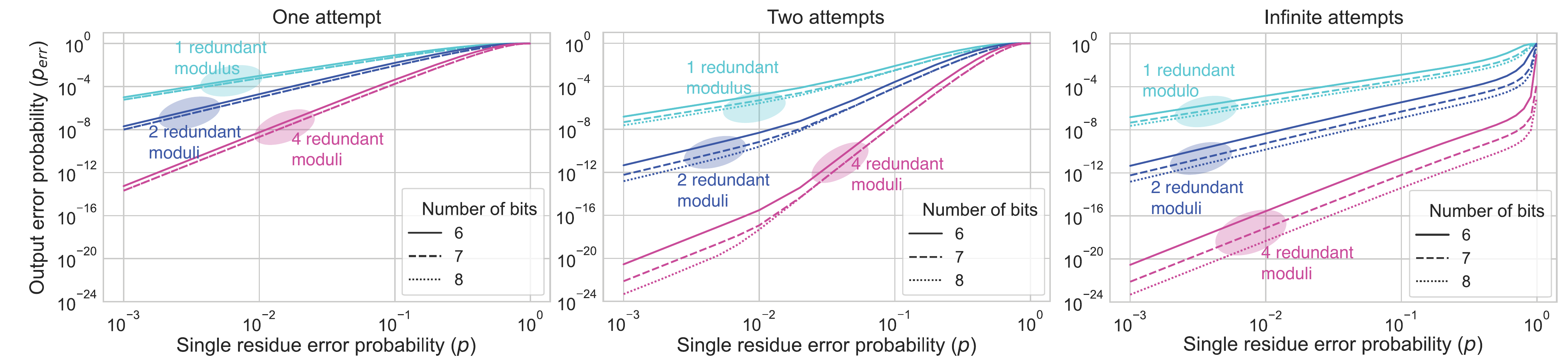}
\caption{Output error probability ($p_{\text{err}}$) for varying number of error correction attempts and number of redundant moduli $(n-k)$ used}
\label{fig:therotical-err}
\end{figure*}

\begin{figure*}[t]
\centering
\includegraphics[width=\textwidth]{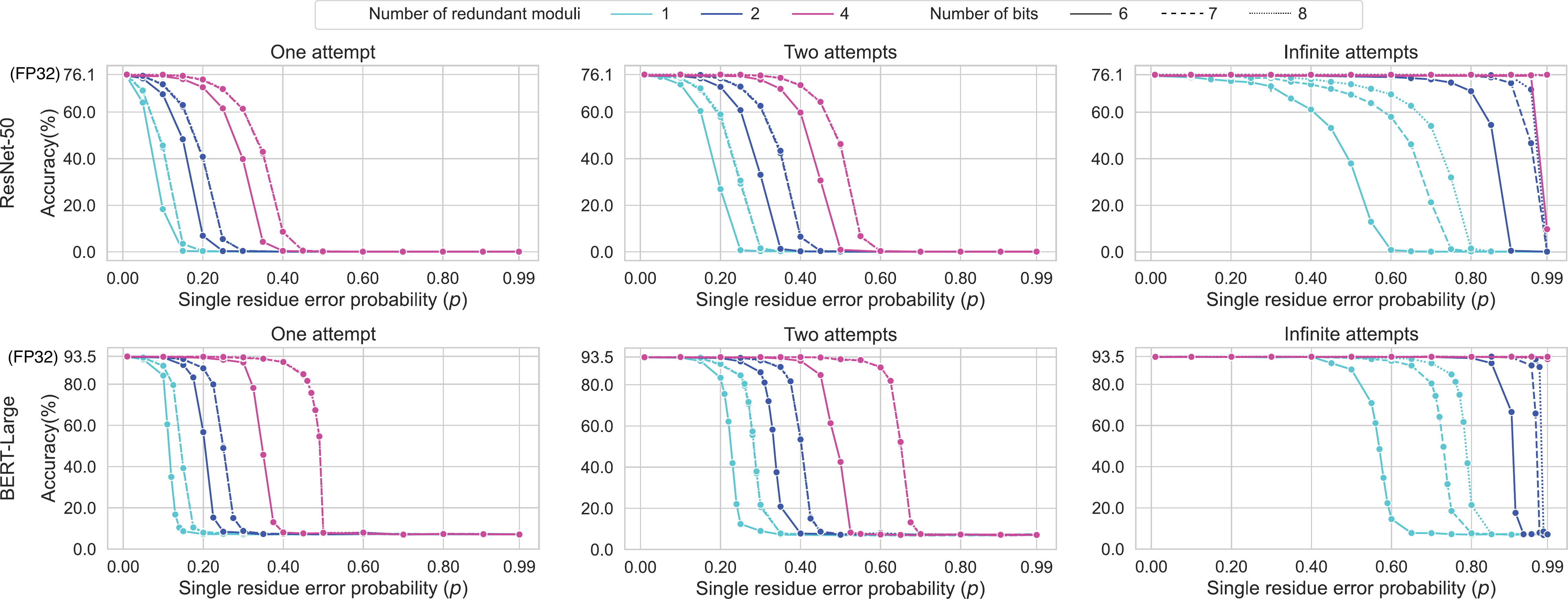}
\caption{ResNet-50 (top row) and BERT-large (bottom row) accuracy for varying number of attempts and number of redundant moduli used.}
\label{fig:exp-acc-err}
\end{figure*}
\ignore{

In the analog GEMM core, we can perform $h\times h$ MAC operations in parallel each cycle. 
Therefore, the larger the $h$ value is, the higher the hardware performance we can achieve. 
In practice, the layer sizes are finite and increasing $h$ after a certain point results in a low utilization and a saturated performance. 
For our analysis, we chose $h$ to be 128 considering the common layer sizes in the evaluated benchmarks, i.e., ??.
The chosen $h$ provides a high parallelism with a high utilization in the GEMM core.
In turn, it requires $\log_2 128=7$ extra bits at the output due to the accumulation of 128 values in each dot product.
It should be noted that this value is a representative number and the methods we provide can be generalized for any tile size $h$. 
Similarly, the moduli sets in Table~\ref{table:moduli-sets} show the largest co-prime numbers within the range.
The moduli sets (i.e., number of moduli and the values of moduli) can be picked differently for minimizing the total hardware cost.
Additionally $b_\text{DAC}$ and $b_\text{ADC}$ can be chosen to be smaller than $b_{in}$ and $b_w$ as long as $M$ is large enough. 
We provide an example in Section~\ref{sec:case-study}.}

\section{Redundant RNS for Fault Tolerance}

Any analog compute core is sensitive to noise.
Furthermore, in the case of RNS, even small errors in the residues result in a large error in the corresponding integer the residues represent.
We use the Redundant Residue Number System (RRNS)~\cite{rrns-codes-hanzo} to perform error detection and correction to improve the fault tolerance of the RNS-based analog core.
RRNS uses a total of $n$ moduli: $k$ non-redundant and $n-k$ redundant. 
An RRNS($n,k$) code can detect up to $n-k+1$ errors and can correct up to $\lfloor \frac{n-k}2 \rfloor$ errors.
In particular, the error in the codeword (i.e., the $n$ residues representing an integer in the RRNS space) can be one of the following cases:
\begin{itemize}
    \item \textbf{Case 1:} No error or correctable error (fewer than $\lfloor \frac{n-k}2 \rfloor$ residues have errors),
    \item \textbf{Case 2:} Detectable but not correctable error (more than $\lfloor \frac{n-k}2 \rfloor$ residues have errors, and the erroneous codeword does not overlap with another codeword in the RRNS space),
    \item \textbf{Case 3:} Undetectable error (more than $n-k+1$ residues have errors, the erroneous codeword overlaps with another codeword in the RRNS space).
\end{itemize}

By using RRNS, errors can be detected by using a voting mechanism wherein we divide the total $n$ output residues into $C_k^n$ groups with $k$ residues per group. 
One simple way of voting is to convert the residues in each group back to the standard representation via CRT to generate an output vector for each group\footnote{Here, using CRT on each group can be too expensive for large number of moduli. Typically, error detection and correction is implemented via more efficient base-extension-based algorithms~\cite{base-ext}.}. 
Then, we compare the results of the $C_k^n$ groups. 
If more than $50\%$ of the groups have same result in the standard representation, then the generated codeword is correct.
This corresponds to \textbf{Case 1}.
In contrast, not having a majority indicates that the generated codeword is erroneous and cannot be corrected. 
This corresponds to \textbf{Case 2}.
In this case, the detected errors can be eliminated by repeating the dot product and voting steps, and recalculating the result. 
In \textbf{Case 3}, there are more than $n-k+1$ erroneous residues and the erroneous codeword generated by majority of the groups overlaps with another codeword. 
As a result, the voting mechanism incorrectly determines that the generated codeword is the correct codeword, i.e., the errors in the residues go undetected.

To understand this better, for a single codeword and a single attempt, assume $p_c$ is the probability that there are no errors or the errors are correctable (\textbf{Case 1}), $p_d$ is the probability that errors are detectable but not correctable (\textbf{Case 2}), and $p_u$ is the probability that the errors are undetectable (\textbf{Case 3}).
In this case, $p_c + p_d + p_u = 1$.
The error probabilities $p_c, p_d$ and $p_u$ for a probability of error in a single residue $(p)$ can be computed using equations formulated by James et al.~\cite{rrns-2015} and Peng et al. \cite{ rrns-codes-hanzo}.\footnote{
The detailed equations not shown in this paper due to space constraints.}
For $R$ repeated attempts of performing a dot product to correct the error, the probability of having an erroneous output codeword ($p_\text{err}$) is 
\begin{equation}
p_\text{err}(R) =  1 - p_c  \sum_{k=1}^{R} (p_{d})^k.
\label{eq:p_err}
\end{equation}
As we increase the number of attempts, the output error probability decreases and converges to: $\lim_{R\to\infty} p_\text{err}(R) = p_u/(p_u+p_c)$.
 
The values of $p_\text{err}$ for different numbers of redundant moduli $(n-k)$, numbers of attempts $(R)$, and moduli sets (with different number of bits) are plotted in Fig.~\ref{fig:therotical-err}. 
Broadly, as the probability of a single residue error $p$ increases, the output error probability tends to $1$.
For a given number of attempts, increasing bit-precision and the number of redundant moduli decreases $p_\text{err}$. 
For a fixed number of redundant moduli and a fixed number of bits per moduli, $p_\text{err}$ decreases as the number of attempts increases. 

We investigated the impact of the noise on the accuracy of the MLPerf benchmarks when using RRNS. 
We observe similar behavior in different networks, thus we only show the accuracy results for ResNet50 and BERT-large in Figure~\ref{fig:exp-acc-err}.
Broadly, adding extra moduli and increasing the number of attempts decrease $p_\text{err}$ and help maintain the accuracy for higher $p$ values. 

ResNet50 requires ${\sim} 3.9$ GigaMAC operations (GOp) for one inference on a single input image. 
For a $128\times128$ MVM unit, the total number of MVM outputs is ${\sim}29.4$M. 
For all output values to be correct, $p_\text{err}\leq1/29.4$M = $3.4\times 10^{-8}$.
This $p_\text{err}$ value needs to be $\leq1/358.6$M = $2.8\times 10^{-9}$ for BERT-large.
However, from Fig.~\ref{fig:exp-acc-err}, we observe that the DNNs are resilient to noise and the tolerable $p_\text{err}$ is much higher than the calculated numbers. 
The accuracy of ResNet50 only starts degrading (below $99\%$ FP32) when $p_{\text{err}} \approx 4.9 \times10^{-5}$ (${\sim}10^3 \times$ higher than the estimated value) on average amongst the experiments shown in Figure~\ref{fig:exp-acc-err}. 
This cut-off probability is $p_{\text{err}} {\approx} 4\times10^{-4}$ for BERT-large (on average ${\sim} 10^5\times$ higher than the estimated value).

\section{Energy Efficiency of RNS-Based Core}

In this section, we show that an RNS-based analog core is more energy efficient than their fixed-point counterparts \emph{for the same precision}.
High-precision data converters---particularly ADCs---dominate the power consumption in analog DNN accelerators~\cite{Kim:18, rekhi-2019}. 
Our RNS approach alleviates this high power consumption of ADCs without compromising accuracy. 
To quantify our findings, we estimate the energy consumption of a DAC per conversion as:
\begin{equation}
        E_{\dac} = \text{ENOB}^2 C_u V_{\text{DD}}^2,
\end{equation}
where $C_u = 0.5$ fF is a typical unit capacitance and $V_{\text{DD}}$ = 1V is the supply voltage~\cite{murmann-mixed-signal}.
The energy consumption of an ADC per conversion is estimated as:
\begin{equation}
        E_{\adc} = k_1\text{ENOB} + k_24^\text{ENOB},
\end{equation}
where $k_1 {\approx}100$ fJ and $k_2 {\approx} 1$ aJ~\cite{murmann-mixed-signal, murmann-adc-survey}.
The exponential term (i.e., $k_2 4^\text{ENOB}$) dominates at large ENOB (after ${\sim}10$-bits).
 
\begin{figure}[t]
\centering
\includegraphics[width=\linewidth]{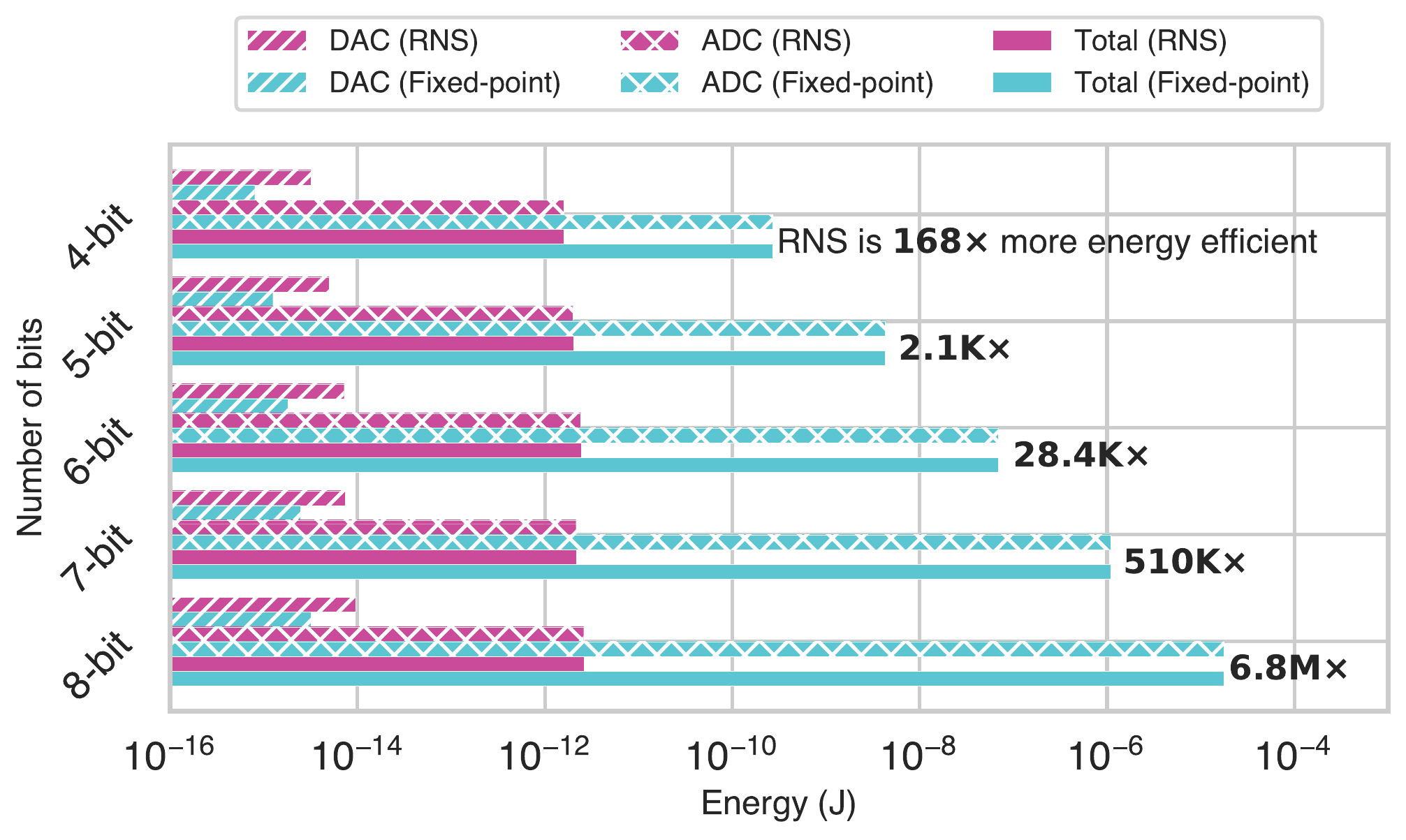}
\caption{Energy consumption of ADCs and DACs per element for RNS-based ($n$ conversions) and fixed-point (1 conversion) analog hardware.
}\label{fig:energy}
\end{figure}

Fig.~\ref{fig:energy} shows total $E_{\dac}$ and $E_{\adc}$ for both the RNS-based and the regular fixed-point analog hardware configurations that were shown previously in Table~\ref{table:moduli-sets}. 
Here, different from Table~\ref{table:moduli-sets}, $b_{\adc}=b_{\out}$ for the regular fixed-point core to achieve the same precision as the RNS approach.
To achieve the same MVM throughput in both cores, the RNS-based core with $n$ moduli must use $n$ MVM units---leading to $n$ DACs and $n$ ADCs.
From Figure~\ref{fig:energy}, we observe that ADCs have approximately three orders of magnitude higher energy-consumption compared to DACs with the same ENOB. 
Energy of an ADC is exponentially dependent on ENOB, and so even with $n$ ADCs in the RNS-based core, its energy consumption is still $168\times$ to $6.8$M$\times$ lower than the regular fixed-point core.

Moreover, besides data converters, RNS reduces the required signal-to-noise ratio (SNR) in the analog MVM units. 
The energy consumption of the analog MVM unit depends on the SNR for the analog signals, and this SNR increases exponentially with the desired compute precision.
Thus, RNS brings additional savings by allowing the MVM units to work with lower SNR in the analog domain.

In the case of using RRNS, the number of DACs, ADCs, and analog MVM units will increase linearly with each added redundant moduli.
Although this increases $n$ and the energy consumption of data converters linearly, considering the high energy efficiency gains (up to six orders of magnitude) over the fixed-point cores, this extra cost for extra moduli is tolerable. 
Here, the number of redundant moduli and number of attempts required to achieve fault-tolerant computing is dependent on the the noise distribution in the chosen technology and the DNN model.  

Different from the regular fixed-point hardware, the use of RNS requires forward and reverse conversion between the RNS and the standard representation, and modulo operation in the analog domain. 
The forward conversion is simply a modulo operation whereas reverse conversion is performed via CRT (See Eq.~\eqref{eq:crt}).
We synthesized the RTL design of these conversion circuits using ASAP 7nm technology library~\cite{Clark2016ASAP7A7}.
The modulo operations are optimized using Barrett Reduction~\cite{barrett-red}.
These converters consume up to ${\leq}0.1$ pJ per conversion (forward and reverse in total), which is negligible.
Analog modulo can be performed in different ways depending on the analog technology used. 
One can use ring oscillators~\cite{analog-modulo-2018} to perform modulo operation electrically. 
This approach uses an odd number of $m$ inverters to oscillate the signal.
The location of the falling/rising signal winds back after each $m$ cycles as the signal oscillates. 
By keeping track of the start and the end location of the signal between a specific time interval proportional to the value $x$, the modulo operation of $x$ against $m$ can be obtained. 
As a set of inverters is a trivial circuitry, the ring oscillator does not add much to the energy consumption. 
Alternatively, one can perform modulo optically by using the phase in the optical domain. 
Addition with phases is effectively a modular addition against $2\pi$ as the phase values wind back at every $2\pi$.
The phase of a signal at the output of a row of $h$ phase shifters will be $|\Sigma_{l=0}^h \phi_l|_{2\pi}$.
Multiplying the values with $2\pi/m$ before the modular addition enables one to perform modulo against arbitrary $m$ instead of only $2\pi$.
Modulo operation can thus be performed along with accumulation without any additional cost.

\section{Conclusion}

In this paper, we present a generic dataflow framework for an RNS-based analog accelerator that is agnostic of the analog technology.
We show that an RNS-based analog core provides $99\%$ FP32 accuracy for state-of-the-art DNNs by using 6-bit arithmetic with RNS.
We provide an error detection and correction method using RRNS that improves the fault-tolerance of the analog core against noise. 
We also show that our RNS-based analog hardware can reduce the data conversion energy by multiple orders of magnitude compared to a fixed-point analog hardware at the same precision.
While we provide a generic perspective, for a desired analog technology, one can explore the various trade-offs discussed in this paper to optimize the accelerator micro-architecture for further energy efficiency.
Overall, we believe that RNS is a promising numeral system for the development of the next-generation energy-efficient fault-tolerant analog accelerators.
\bibliographystyle{ieeetr}
\bibliography{rns-dac-2023.bib}

\end{document}